\documentclass[%
 reprint,
 amsmath,amssymb,
 aps,
]{revtex4-2}

\usepackage{graphicx}
\usepackage{dcolumn}
\usepackage{bm}
\usepackage{xcolor}
\usepackage{amsmath}
\usepackage{amssymb}
\usepackage{color}

\begin{document}

\preprint{APS/123-QED}

\title{Optimizing Population Accumulation in Quantum States Using Microwave Spectroscopy}

\author{Jia-You Liou}
\author{Chi-En Wu}%
\author{Hsuan-Jui Su}%
\author{Yi-Hsin Chen}
\altaffiliation[Also at ]{Center for Quantum Technology, Hsinchu 30013, Taiwan}
\email{yihsin.chen@mail.nsysu.edu.tw}
\affiliation{%
Department of Physics, National Sun Yat-Sen University, Kaohsiung 80424, Taiwan
 }%

\date{\today}

\begin{abstract}
We present an all-optical method for efficiently preparing cold atoms in a desired Zeeman state, either on the magnetically insensitive clock state ($m_F=0$) or a particular state suitable for processing or storing quantum information. By applying the theoretical fitting model to a single microwave spectrum, we can individually determine the population distribution, microwave polarization ratio, and microwave Rabi frequency. We can dynamically track the population distribution during the optical pumping process using this real-time microwave spectrum. In a steady-state condition, a simplified model, which considers resonant and off-resonant transitions, indicates that there is an upper limit to the purity under a weak optical pumping field. The population purity up to $96(2)\%$ or $98(1)\%$ on the desired quantum state has been achieved after optimizing the intensity and polarization of the optical pumping field. Our study provides valuable information and potential applications in precision measurement and quantum computation research.
\end{abstract}

\maketitle

\newcommand{\FigOne}{
\begin{figure}[t]
	\includegraphics[width=\columnwidth]{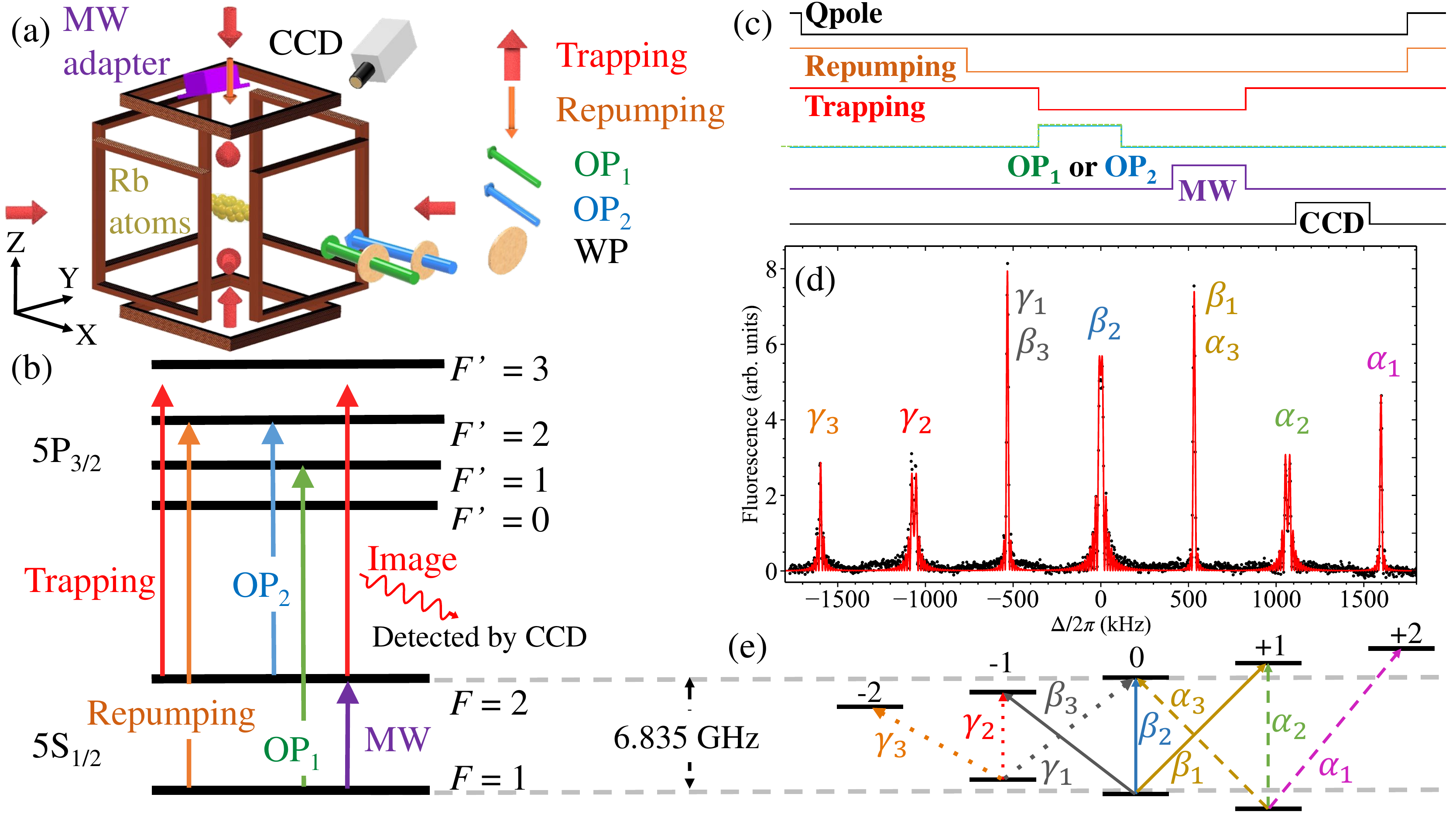}
	\caption{(a) Schematic of the experimental setup. The Qpole field used for capturing the cold atoms is not shown in the figure. WP: waveplate. (b) Relevant energy levels for $^{87}$Rb atoms and laser excitations. Optical pumping fields OP$_1$ and OP$_2$ were applied to pump the population away from hyperfine states $F=1$ and $F=2$, respectively. The trapping field, which captures atoms, also serves as an image field. A mono-color CCD camera  was employed to detect the fluorescence signals. A microwave field was frequency scanned around 6.835 GHz to obtain the microwave spectrum. (c) Timing sequence for the cooling process, population preparation, and population detection. (d) A typical microwave spectrum displays seven major peaks under a magnetic field in an arbitrary direction. (e) Microwave transitions between the Zeeman sublevels. The energy splittings for $F=2$ and $F=1$ are $m_F\times$0.7 MHz/G and $m_F\times$(-0.7) MHz/G, respectively.
}
	\label{fig:Exp Scheme}
	\end{figure}
}	
\newcommand{\FigTwo}{
\begin{figure}[t]
	\includegraphics[width=1\columnwidth]{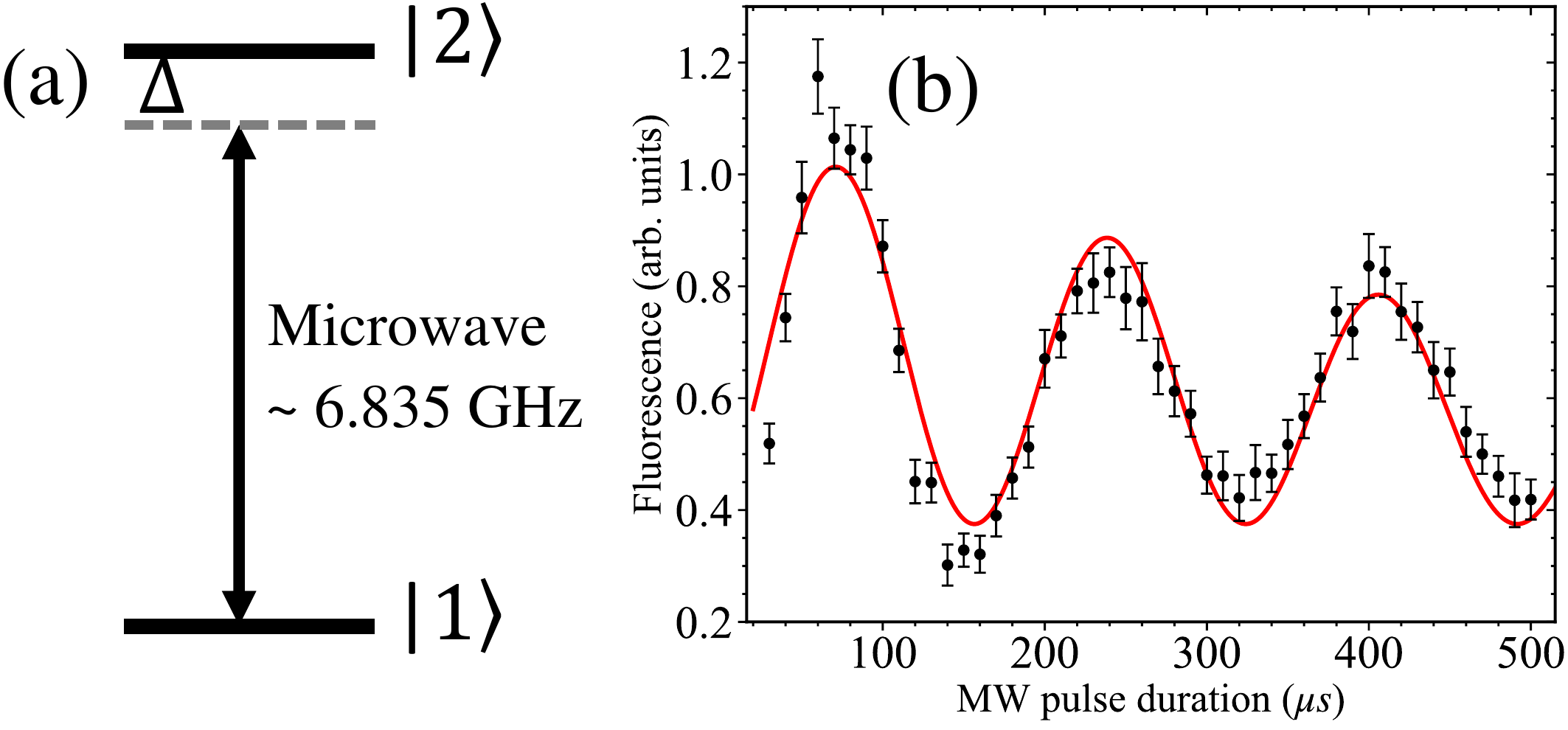}
	\caption{(a) Energy structure of a two-level system. The microwave field is tuned to the frequency of the transition at approximately 6.835 GHz with a certain detuning $\Delta$. (b) Directly determine the Rabi frequency from the specific transition, e.g., $\alpha_1$ transition $|F=1, m_F=1 \rangle \rightarrow |F=2, m_F=2 \rangle$, which gives the Rabi frequency of $2\pi \times $6.0 kHz. 
	}
	\label{fig:Rabi}
	\end{figure}
}
\newcommand{\FigThree}{
\begin{figure}[t]
	\includegraphics[width=\columnwidth]{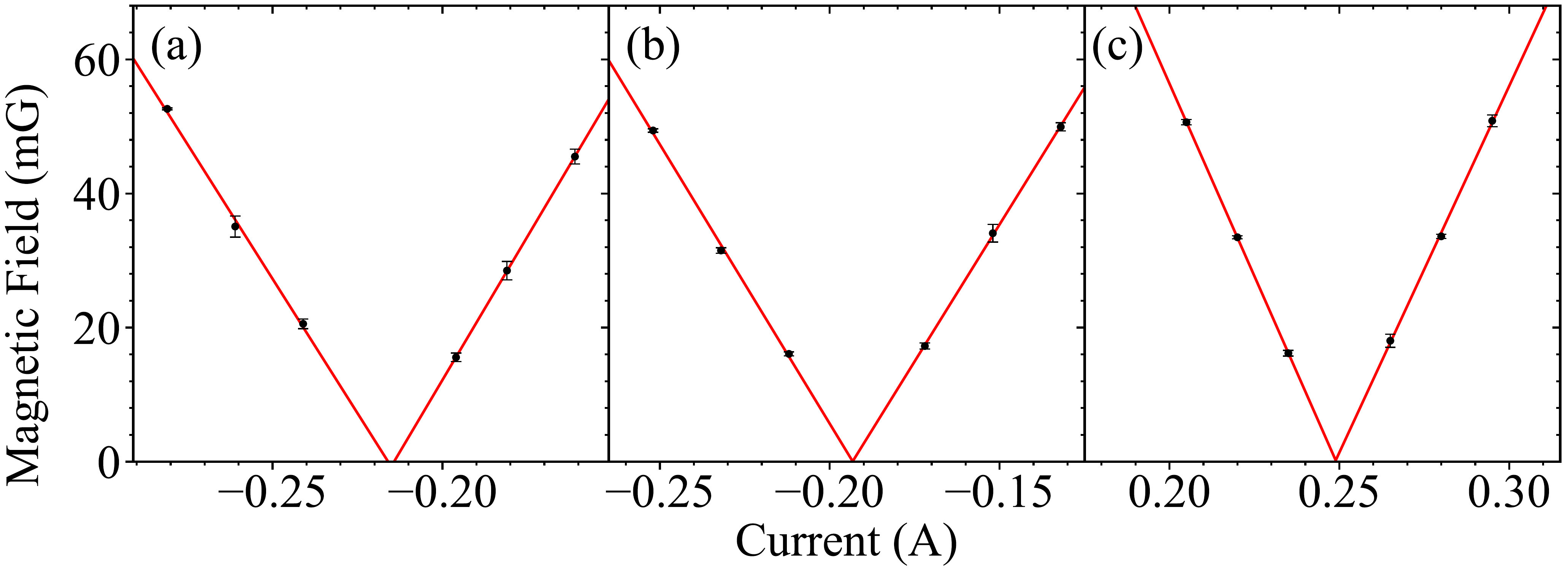}
	\caption{The magnetic field strength derived from the Zeeman splittings as a function of the applied coil current in X (Fig.~a), Y (b), and Z (c) axes after minimizing the stray field. The red linear fitting lines give that the DC magnetic fields, $B_{coil}$, were 0.80, 0.83, 1.1 Gauss/Amp. By extrapolating both linear fits to the magnetic field, we derive the resolution to minimize the magnetic field below two mG.  
}
	\label{fig:Bfield}
	\end{figure}
}
\newcommand{\FigFour}{
	\begin{figure}[t]
	\includegraphics[width=0.95\columnwidth]{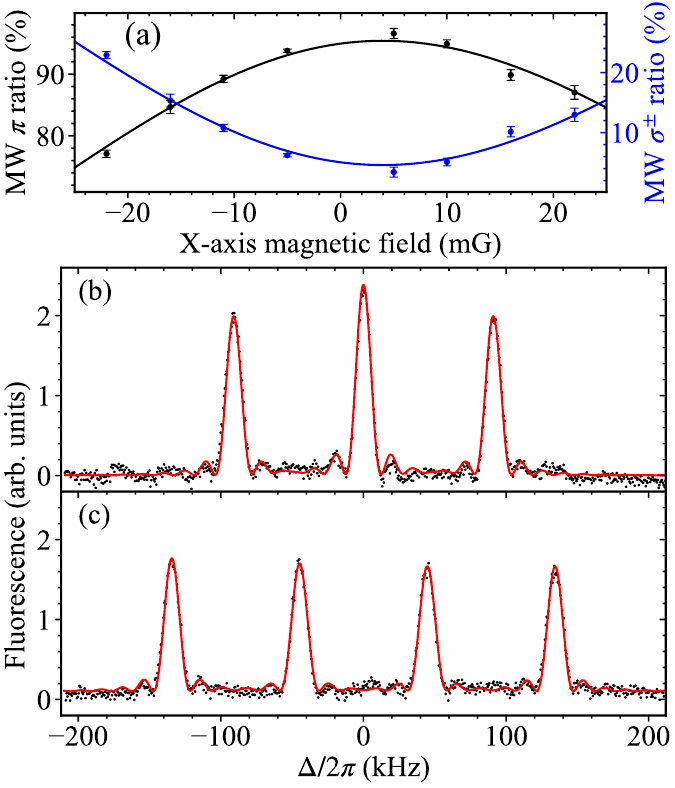}
	\caption{(a) The microwave polarization measurements as a function of the magnetic field strength on the X axis under an additional magnetic field of 65 mG on the Y axis, which had the same direction as the MW polarization. A 97$\%$ polarization purity for $\pi$ transition has been achieved. (b) and (c) are the spectral measurements when the external magnetic field was applied on the Y and Z axes, respectively. The solid lines are the best fits by applying Eq.~(\ref{eq:Rabi}) and considering all possible MW transitions. The best fits show that the population was equally distributed in the present measurements.} 
	\label{fig:MWpolarization}
	\end{figure}
}

\newcommand{\FigFive}{
	\begin{figure}[t]
	\includegraphics[width=1\columnwidth]{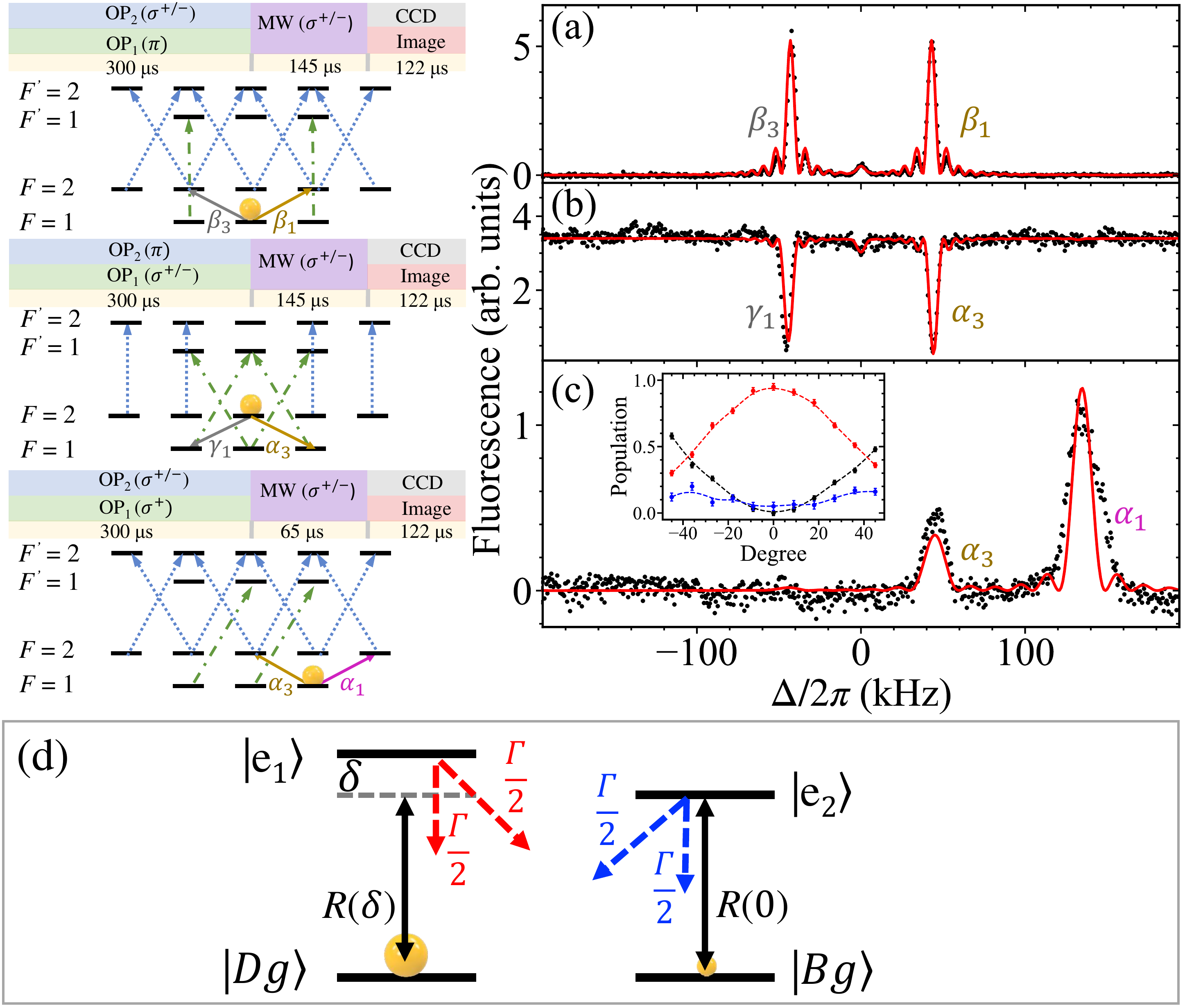}
	\caption{Timing sequence, the relevant energy levels, and laser excitations. (a), (b), and (c) are the microwave spectra for atomic preparation in states $|F=1,m_F=0\rangle$, $|2, 0\rangle$, and $|1,1\rangle$, respectively. The best purity of population in these states were $98(1)\%$, $98(2)\%$, and $96(2)\%$, from (a) to (c). The red, black, and blue circles in the inset represent the populations $P_\alpha$, $P_\beta$, and $P_\gamma$ as a function of the orientation angle of a zero-order $\lambda/4$ plate for OP$_1$ beam. (d) A simplified set of atomic energy levels involving one dark state $|Dg\rangle$, one bright state $|Bg\rangle$, and two excited states, $|e_{1(2)}\rangle$. These states are coupled by a resonant $R(0)$ and a far-off resonant $R(\delta)$ driving fields. 
}
	\label{fig:peakanddip}
	\end{figure}
}

\newcommand{\FigSix}{
	\begin{figure}[t]
	\includegraphics[width=1\columnwidth]{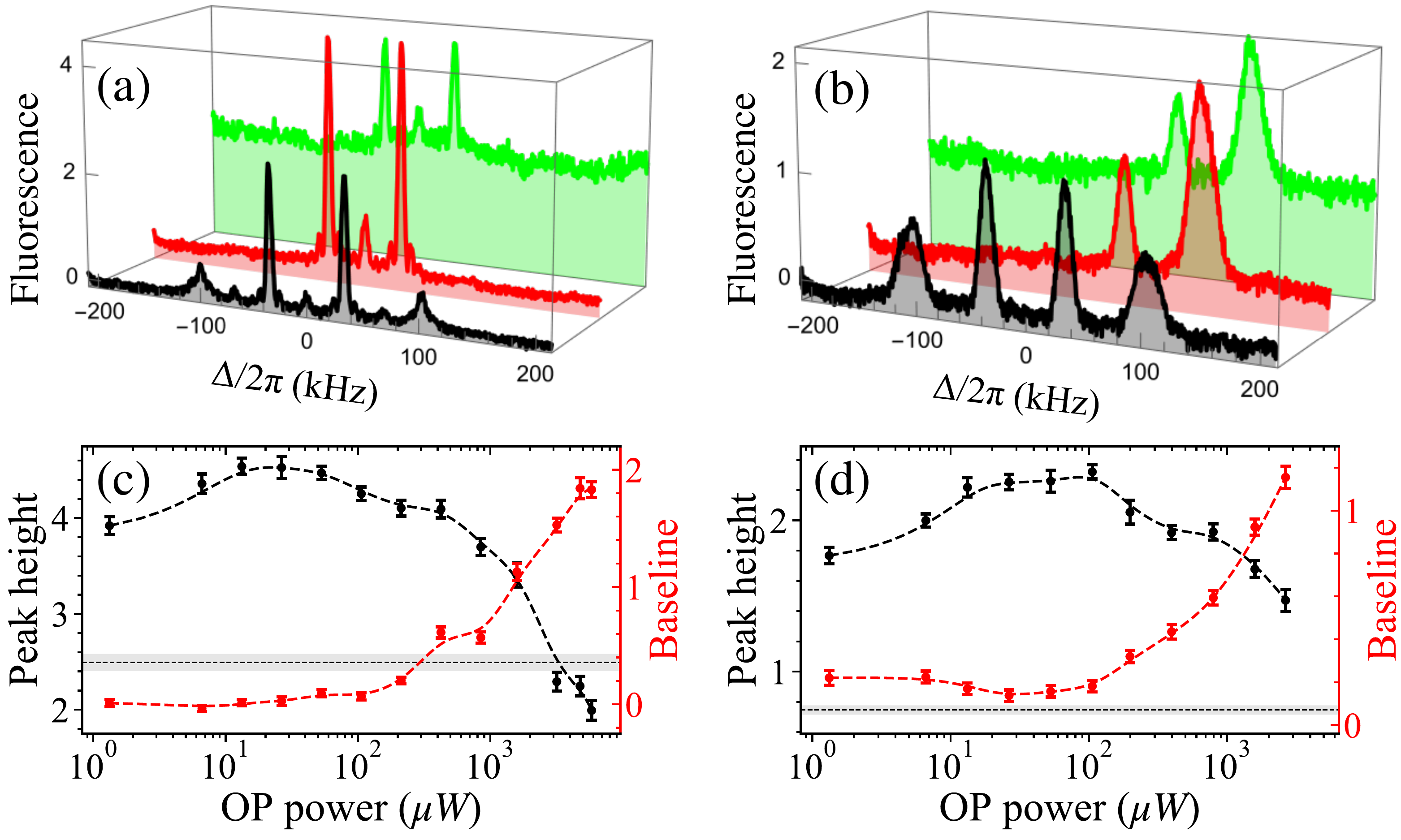}
	\caption{(a) and (b) illustrate the fluorescence signals for identifying the optimal population accumulation in states $|1,0\rangle$ and $|1,1\rangle$, respectively. The black line represents signals without the OP field, the red one with an optimized OP field, and the green one with the strongest OP field. We extract the peak heights ($\beta_1$ and $\alpha_1$) and baseline levels in (c) and (d). The dashed lines are the guide to the eyes. The dotted lines are the extracted peak heights without the OP field. Vertical axes are displayed in arbitrary units. Both measurements show that the population $P_\beta$ in (c) or $P_\alpha$ in (d) was enhanced with increasing pumping power; however, further increasing the power reduced the population. Meanwhile, the baseline level increased with increasing pumping power, so more populations were pumped to another hyperfine state $F=2$ via off-resonant transitions. The fluorescence signal for the $\alpha_1$ transition has improved by a factor of 3, corresponding to the 3-fold population increment. Therefore, the population $P_\beta$ in (c) or $P_\alpha$ in (d) can be optimized by applying a proper OP power (or intensity). 
 }
	\label{fig:OPparameters}
	\end{figure}
}
	
\section{\label{sec:introduction}Introduction}
Preparing atomic populations into a specific Zeeman state is essential for quantum information science and precision measurements, such as quantum memory~\cite{YHC2013memory,Cho2016memory,Hsiao2018memory,Laurat2018}, quantum manipulation~\cite{YHC2012,Chang2021}, single-photon generation~\cite{Martin2013}, atomic magnetometry~\cite{RMP2002,Chalupczak2010,Thiele2018}, and atomic clock~\cite{DiPRA2010,Hamzeloui2016,Wang2018,Fedorova2020}. 
Accumulating the population in the desired Zeeman sublevel can increase the optical density (OD) of the atomic medium according to the Clebsch-Gordan coefficient of the selected atomic transition, and can avoid energy loss in the manipulation of the light retrieval~\cite{Guan2007,Hsiao2014,Hsiao2018memory,Laurat2018}.
To make two light fields strongly interact, the researchers prepared the population in the two specific ground Zeeman states, forming two motionless light pulses via the quantum interference~\cite{YHC2012}. A phase shift of $\pi$ by a single photon has been proposed in such an ultrahigh OD system~\cite{YHC2021JOSAB}. 
In addition, increasing the population in the clock state ($m_F=0$), which is a magnetically insensitive state, leads to an apparent reduction in the noise for the atomic clock~\cite{Wang2018}.
Optical pumping (OP) is a standard method to pump all atoms into an uncoupled dark state.
By applying the stimulated Raman adiabatic passage (STIRAP), one can coherently transfer the population between two quantum states via two coherent light pulses~\cite{Du2014}. Moreover, through combining optical pumping with microwave pumping, the researchers prepared the population in a clock state with a purity of $83\%$~\cite{Wang2018}. With appropriate polarization configuration and laser power, state purities of more than $96\%$~\cite{Wang2007} or $97\%$~\cite{Fang2011} were achieved. 

\FigOne
At present, the methods to prepare and detect the population distribution include measuring (1) the electromagnetically-induced-transparency-based transmission~\cite{Wang2007}, (2) the transition between different Zeeman states with the same quantum number $F$~\cite{Chalupczak2010,Chalupczak2012}, (3) the hyperfine state transition ($\Delta F=\pm 1$) driving by a microwave~\cite{Micalizio2009,DiPRA2010,Szymaniec2013,Sinuco2019,Wei2020,Bouton2021}, and (4) the Ramsey atom interferometer~\cite{Wu2013,Wang2018}. In addition, the atomic population distribution can be detected via the reflected microwave power spectrum using a real-time nondestructive imaging system~\cite{Dubosclard2021}. 
In this work, we applied microwave spectroscopy to measure the population distribution and provided an analytical solution for atomic state detection. The Rabi frequency of the applied microwave was individually determined from the detection of the population cycling transition. 
Microwave pulses with high accuracy would be helpful in the atomic ground-state manipulation and for the Ramsey interference study~\cite{Clemmen2016,Wu2013,Wang2018,Wang2020,Niemietz2021}.
The environment magnetic field can be compensated with a precision of below two mG from microwave spectroscopy. After knowing the population distribution, the optical pumping method was employed to pump the population to the desired quantum state with high purity up to $96(2)\%$ or $98(1)\%$. An estimation based on the far-off-resonant transitions, which involves pumping the atoms away from the dark state, shows an upper limit of $99.8\%$. Experimentally, the impure polarization of light and the non-uniform magnetic field affect the state purity.
The main feature and time evolution during the OP of microwave spectroscopy will be discussed and compared with our theoretical predictions. 

\section{Scheme and experimental setup}
We perform the experiment using $^{87}$Rb atoms in a typical magneto-optical-trap (MOT) system, including three pairs of trapping beams and one repumping beam. The magnetic field gradients in the X, Y, and Z directions of 3.06, 1.44, and 4.97 Gauss/cm were produced by a pair of anti-Helmholtz coils (Qpole) with a current of 2.0 Amp.  
The trapped atom cloud has a dimension of $7.5\times 5.7 \times 3.7$ mm$^3$ containing $2\times 10^8$ atoms. The schematic of the setup, relevant energy levels, and laser excitations are shown in Figs.~\ref{fig:Exp Scheme}(a) and \ref{fig:Exp Scheme}(b).
The population can be prepared by trapping, repumping, and optical pumping fields. The pumping fields OP$_1$ and OP$_2$ drove the population away from $F=1$ and $F=2$, respectively. Figure \ref{fig:Exp Scheme}(c) shows the timing sequence. The repetition time was 436 ms, including 430 ms for the cooling process and 6 ms for preparing and detecting the population. 
Denote the time of switching on the microwave as $t=0$. We typically turned off the Qpole field, repumping, and trapping beams of the MOT at $t=-6$~ms, $-1.12$~ms, and $-0.52$~ms in sequence. 
In order to characterize the state-preparation efficiency, we use microwave spectroscopy to measure the population distribution on the Zeeman sublevels of the ground state $F = 1$, as shown in Fig.~\ref{fig:Exp Scheme}(d). The microwave lasted for 80 $\mu$s, but the period was varied in other measurements. When turning off the microwave, the trapping beams were turned on again to serve as the image beams. The fluorescence signals were detected by a mono-color CCD camera (The Imaging Source: DMK 22BUC03) with an exposure time of 122 $\mu$s.

After the cooling process, the trapping fields were applied to pump the cold atoms to three Zeeman states of $F=1$. Under a magnetic field, the energy splittings of Zeeman sublevels for $F=2$ and $F=1$ states are $m_F\times$0.7 MHz/G and $m_F\times$(-0.7) MHz/G, respectively. Choose the direction of the magnetic field as the quantization axis. The polarization of the microwave was random in an arbitrary magnetic field. As shown in Fig.~\ref{fig:Exp Scheme}(e), the microwave can induce $\alpha_i$, $\beta_i$, and $\gamma_i$ transitions involving populations in states $\alpha$, $\beta$, and $\gamma$, respectively, where $i = 1, 2, 3$ corresponds to $\sigma^+$, $\pi$, $\sigma^-$ transitions. Therefore, a typical microwave spectrum displays seven major peaks, among which $\alpha_3$ and $\beta_1$ (as well as $\beta_3$ and $\gamma_1$) exhibit the same resonance frequency. 
The 6.835 GHz microwave was generated by a synthesizer (Rigol DSG830) combined with a frequency multiplier (Minicircuits: ZX90-3-812-S+), power amplifier (Minicircuits: ZVE-3W-183+), and waveguide adapter (Woken: 0060WA2KOBB01X). The microwave was scanned at a speed of 0.6 kHz per repetition time.  

\FigTwo

\section{Theoretical model}
We theoretically explain the main feature of the microwave (MW) spectrum and individually determine the population distribution, polarization, and MW Rabi frequency. First, we measured the MW strength. Consider a two-level system with states $|1\rangle$ and $|2\rangle$, and assume the population is initially in $|1\rangle$. The population time evolution is a function of the MW detuning $\Delta$, MW period $t$, and MW Rabi frequency $\Omega$. The population in state $|2\rangle$, $P_2$, is given by
\begin{equation}
P_2(t, \Delta) =  \\ 
\frac{\Omega ^2}{\Omega'^2} \left(1-\text{Cos} (\Omega' t )\right),
\label{eq:Rabi}
\end{equation}
where
\begin{equation}
 \Omega'= \sqrt{\Omega ^2+\Delta^2}. \nonumber \\ 
\end{equation}
The MW Rabi frequency can be estimated by varying the MW transition period and tracking the population oscillation. The MW frequency was fixed for $\alpha_1$ transition, and the fluorescence signals are shown in Fig.~\ref{fig:Rabi}(b). In this measurement, states $|1\rangle$ and $|2\rangle$ represent $|F=1, m_F=1 \rangle$ and $|F=2, m_F=2 \rangle$, respectively. Considering the population decay, the measured oscillation gives the Rabi frequency for the $\alpha_1$ transition of $2\pi\times 6.0$ kHz.

Below we show the procedure to extract all of the unknown parameters from a seven peaks microwave spectrum (e.g., Fig.~\ref{fig:Exp Scheme}(d)), including the population distribution $P_\alpha$, $P_\beta$, $P_\gamma$, the polarization component of the microwave, and the Rabi frequency for each transition. Here we define $\Omega_1$ as the MW Rabi frequency for the dipole matrix element between two states of one. 
The dipole matrix elements of $\sigma^+$ ($\sigma^-$) transitions are ($C_{\gamma_1(\alpha_3)}$, $C_{\beta_1 (\beta_3)}$, $C_{\alpha_1(\gamma_3)}$) = ($\sqrt{1/12}$, $\sqrt{1/4}$, $\sqrt{1/2}$). 
The ones for $\pi$ transitions are $(C_{\gamma_2}, C_{\beta_2}, C_{\alpha_2}) = (\sqrt{1/4}, \sqrt{1/3}, \sqrt{1/4})$ for $\gamma_2$, $\beta_2$, and $\alpha_2$ transitions, respectively. 
We can derive the bias magnetic field from the peak splittings. The MW resonant frequencies were shifted due to the Zeeman effect, and therefore, the detuning $\Delta$ in Eq.~(\ref{eq:Rabi}) needs to be replaced by $\Delta-\Delta_0$, where $\Delta_0$ is the MW resonant frequency. For the spectrum shown in Fig.~\ref{fig:Exp Scheme}(d), $\Delta_0$ was $\pm$ $j\times$ 532.5 kHz for $j=1,2,3$. 

To determine the population distribution, $(P_{\gamma}, P_{\beta}, P_{\alpha})$, we first focus on the $\pi$ transition peaks, i.e., the second, fourth, and sixth peaks. The fluorescence signals are proportional to the total number of atoms and the sensitivity of the CCD camera, which have an equal impact on each peak. The factors $(P_{\gamma}, P_{\beta}, P_{\alpha})$, where $P_{\gamma}+P_{\beta}+P_{\alpha}=1$, can vary the fluorescence peak height and need to be taken into account in Eq.~(\ref{eq:Rabi}).
Here, $\Omega$ in Eq.~(\ref{eq:Rabi}) is replaced by $\Omega_1\times$($C_{\gamma_2}$, $C_{\beta_2}$, $C_{\alpha_2}$). The population distribution was derived from these three peaks of seven, showing that the population was almost equally distributed. 
We next determine the polarization component of the microwave from the most-left ($\sigma^-$-polarization) and the most-right ($\sigma^+$- polarization) peaks. 
The quantization axis is aligned with the applied magnetic field, allowing the MW to drive $\sigma^+$, $\sigma^-$, and $\pi$ transitions. From the comparison of the determined Rabi frequencies that fit the data and the derived ones from the dipole matrix element, we obtain the ratios of polarization components are ($14\%, 80\%, 5.4\%$) for ($\sigma^-$, $\pi$, $\sigma^+$) polarizations. 
In order to verify the accuracy of the above-determined parameters, the third and the fifth peaks were simulated with the above-given parameters. The feature of the spectrum fitted quite well, including the relevant peak height and the oscillation structure.

\FigThree

\section{Results}
We now discuss how to improve the purity of the population accumulation by applying optical pumping. First, if the laser polarization $\hat{Z}_L$ differs from the magnetic field direction $\hat{Z}_B$, the laser field can drive $\Delta m_F = \pm 1$ or  $\Delta m_F=0$ transitions, causing the pumping process to be more complicated. The effective magnetic field $\bf{B_{eff}}$ = $\bf{B_{coil}}+\bf{B_0}$, where $\bf{B_0}$ is the stray field from the environment and $\bf{B_{coil}}$ can be adjusted by the three pairs of the Helmholtz coils. 
Since the Zeeman splitting is proportional to the strength of $\bf{B_{eff}}$, we minimized the stray magnetic field on the three orthogonal directions by measuring the frequency shift for various $\bf{B_{eff}}$. Then, we can define the quantization axis on the $\hat{Z}_B$ direction and purify the optical pumping field transition. 
The frequency splitting for two nearby peaks was obtained by extracting the detuning $\Delta_0$ for several magnetic field strengths along a fixed direction. The spectrum's peak resolution is better than 5 kHz, determined by the full width at half maximum of the peak.
Figures~\ref{fig:Bfield}(a), \ref{fig:Bfield}(b), and \ref{fig:Bfield}(c) show the extracted frequency splittings as a function of the magnetic field on X, Y, and Z axes, respectively, when the stray magnetic field was almost canceled. 
The red fitting lines (either a linear or a quadratic function) show that the DC magnetic fields, $B_{coil}$, were 0.80, 0.83, 1.11 Gauss/A on the X, Y, and Z axes. 
In addition, the intercept points or the minimum points of the fitting lines give the best setting current to compensate the $\bf{B_0}$ and the lowest achievable magnetic field with a resolution of below two mG for each axis. 

\FigFour
Once the stray magnetic field had been compensated, we applied an additional magnetic field as the quantization axis. The polarization components of the optical pumping and microwave fields will be varied according to the direction of the quantization axis. The magnetic field of 65 mG was set on the Y axis, and another minor field on the X axis was varied to determine polarization purity. Based on theoretical fits to the MW spectra, we can determine the purity of the MW polarization. A purity of 97$\%$ for the $\pi$ transition has been achieved, as demonstrated in the corresponding microwave spectrum shown in Fig.~\ref{fig:MWpolarization}(b).
When we set the magnetic field of 65 mG on the X axis, the microwave polarization was perpendicular to this quantization axis, and it drove the $\sigma^+$ and $\sigma^-$ transitions. The four peaks in Fig.~\ref{fig:MWpolarization}(c) correspond to $\gamma_3$, $\gamma_1 + \beta_3$, $\beta_1 + \alpha_3$, and $\alpha_1$ transitions, from the left to the right peaks. According to the best fits to the data, the red lines show that the population was equally distributed in the present measurements. The polarization purity, period, and microwave power are crucial for manipulating $\pi$-pulse transition, such as the Ramsey interference study. 

\FigFive
We then applied optical pumping to accumulate the populations to a Zeeman state and improve state purity. The optical pumping fields, OP$_1$ and OP$_2$, were set at the resonant frequencies of $F=1\rightarrow F'=1$ and $F=2\rightarrow F'=2$ transitions, respectively. 
First, the target state is the clock state $|F=1,m_F=0\rangle$. We applied the external magnetic field on the Z axis as the quantization axis. The polarizations of the MW, OP$_1$, and OP$_2$ fields were $\sigma^\pm$, $\pi$, and $\sigma^\pm$, respectively.
Initially, the population was almost equally distributed among the three Zeeman states. When the trapping beams were turned off at $t=-330 ~\mu s$, OP$_1$ and OP$_2$ were turned on simultaneously and lasted for 300 $\mu$s. As the shown energy levels and laser excitations in Fig.~\ref{fig:peakanddip}(a), the dark state is $|F=1,m_F=0\rangle$. After optimizing the laser polarization and intensity, the MW spectrum displayed only two major peaks corresponding to $\beta_1$ and $\beta_3$ transitions. The best fit shows that the fraction of atoms on this clock state increases to $98(1)\%$. Note that the small peak at the center, with $\Delta=0$, corresponds to a $\pi$-polarized MW transition, with only $3\%$ polarization impurity.
Similarly, we can pump the population in another clock state $|F=2,m_F=0\rangle$ by changing the polarization of the pumping fields, e.g., $\pi$ polarization for OP$_1$ and $\sigma^\pm$ polarization for OP$_2$. Here we observed the dips instead of the peaks. The baseline signals of the fluorescence showed the number of atoms in $F=2$ states. The fluorescence signals were taken using an image beam ($F=2\rightarrow F=3'$ transition frequency with a specified detuning) after the population was initially prepared in the $F=2$ state. However, the resonant microwave drove some of the population to the $F=1$ state, thereby reducing the strength of the resonant fluorescence.

Finally, we accumulated the population in the state $|F=1,m_F=1\rangle$ as it exhibits the strongest coupling strength between two hyperfine states, resulting in the highest optical density. 
The magnetic field was set on the X axis, while OP$_1$ with $\sigma^+$ polarization and OP$_2$ with $\sigma^\pm$ polarization were utilized.
The MW duration was adjusted based on the coupling strength to achieve a $\pi$-pulse transition for the $\alpha_1$ peak. Note that the linewidth of the $\alpha_1$ peak was broader than the fitting curve because of the non-uniform magnetic field across the center and the edge of the atomic cloud~\cite{Masterson1993}. This broadening effect was observed only when the magnetic field was aligned along the X axis. A longer atomic cloud was observed on the X axis relative to the Z axis, making the magnetic field's homogeneity crucial. To optimize the purity of each desired Zeeman state, we well adjusted the orientation of a zero-order $\lambda/4$ plate for the OP$_1$ (or OP$_2$) beam. Only the measurements for state $|1,1\rangle$ are shown in the inset of Fig.~\ref{fig:peakanddip}(c). The polarization impurity causes population to be pumped to the bright states, resulting in clear peaks on the spectrum which are attributed to the bright-state atoms. The purity has been optimized up to $96(2)\%$. Compared to the purity of states $|1,0\rangle$ and $|2,0\rangle$, the lower purity and larger error in state $|1,1\rangle$ are mainly due to the wider transition peaks caused by the non-uniform magnetic field.

Off-resonant transitions hinder the population purity despite having only one dark state in the optical pumping process. We consider a simplified set of atomic energy levels, as depicted in Fig.~\ref{fig:peakanddip}(d). The populations in the dark state ($|Dg\rangle$) and bright state ($|Bg\rangle$) are $N_0(t)$ and $N_1(t)$, respectively. 
We assume the branching ratios of the radiative decay from the excited state to the ground states are equal, i.e., $\Gamma/2$ ($\Gamma$ is the total nature decay rate of the excited state). The absorption rate $R$ depends on the detuning and laser intensity~\cite{OPsimplified2017},
\begin{equation}
R(\delta) =  C_{ij} 
\frac{\Gamma}{2}\frac{I/I_{sat}}{1+4(\delta/\Gamma)^2+(I/I_{sat})},
\label{eq:absorption rate}
\end{equation}
where $I$ and $I_{sat}$ are the laser intensity and saturation intensity, and $\delta$ is the detuning of the laser field from the atomic resonance, $C_{ij}$ denotes the Clebsch-Gordan coefficient involving the transition.
The population transfer between the dark and bright states involves two steps: (1) atoms are excited either from the dark state $|Dg\rangle$ to the off-resonant excited state $|e_1\rangle$ or from the bright state $|Bg\rangle$ to the resonant excited state $|e_2\rangle$, and (2) they then spontaneously decay either from $|e_1\rangle$ to $|Bg\rangle$ or from $|e_2\rangle$ to $|Dg\rangle$. The time evolution of the population can be described as
\begin{equation}
\frac{dN_0(t)}{dt}= \frac{1}{2} N_1(t)R(0)-\frac{1}{2}N_0(t)R(\delta).
\label{eq:population}
\end{equation}
In the steady-state condition, the population purity in the dark state is 
\begin{equation}
\frac{N_0}{N_1+N_0}=\frac{R(0)}{R(0)+R(\delta)}.
\label{eq:poratio}
\end{equation}
Therefore, consider the detunings of the near hyperfine levels of the $^{87}\rm{Rb}$ $D_2$-line transitions, $\delta=12\Gamma$ and $24\Gamma$. We estimate the upper limit of the purity approach $99.8\%$ at low intensities, where $I\ll I_{sat}$. The limitation can be further improved by selecting the $^{87}\rm{Rb}$ $D_1$ transitions, which have a more significant splitting between the two hyperfine levels.

\FigSix
As found in Eqs.~(\ref{eq:absorption rate})--(\ref{eq:poratio}), the pumping field intensity is another factor influencing purity. The duration of the optical pumping process plays a similar role in the population accumulation that we did not discuss here. Figure~\ref{fig:OPparameters} (a) illustrates the fluorescence signals for identifying the optimal population accumulation in state $|1,0\rangle$. The signal in the absence of an optimized OP field is represented by the black line, while the red line represents the signal with an optimized OP field and the green line represents the signal with the strongest OP field. 
A stronger fluorescence signal indicates a larger population. By increasing the intensity of the optical pumping, more population is transferred to the hyperfine state with $F=2$ through off-resonant transitions. As a result, the baseline level of the MW spectrum increases.
We then extract the peak heights ($\beta_1$ transitions) and baseline levels from the spectrum as a function of the OP power, as shown in Fig.~\ref{fig:OPparameters}(c). The black dotted line is the peak height without the OP field. Population (black circles) can be concentrated in a specific state with proper pumping power. The purity of $98(1)\%$ has been achieved from the systematic measurements. Similarly, results for state $|1, 1\rangle$ are shown in Figs.~\ref{fig:OPparameters}(b) and \ref{fig:OPparameters}(d). Both figures demonstrate that the population $P_\beta$ in (c) or $P_\alpha$ in (d) increases with higher OP power. However, increasing the power beyond a certain point results in a decrease in population. The fluorescence signal of the $\alpha_1$ transition has been improved by a factor of three, which corresponds to a three-fold increase in population. In this case, population accumulation was optimized to $98(1)\%$ in the $|1,0\rangle$ state and $96(2)\%$ in the $|1,1\rangle$ state, with limitations imposed by the inhomogeneous magnetic field and far-off-resonant transitions. 
Once the population is accumulated in a single Zeeman state, people can apply the stimulated-Raman adiabatic passage (STIRAP) procedure, which is a two-photon adiabatic transition in a three-level system, to pump the population to any other desired state~\cite{Du2014,Shore2017}. 

\section{Conclusions}
We present a comprehensive approach to population accumulation on any desired Zeeman states. Using a theoretical fitting model, we can individually determine the population distribution, the MW polarization ratio, and the MW Rabi frequency by applying it to a single real-time MW spectrum under an arbitrary magnetic field. The environment's stray magnetic field can be compensated below two mG from the MW spectroscopy. The state purities were optimized by adjusting OP polarization, period, and intensity. In a steady-state condition, a simplified model, which considers resonant and off-resonant transitions, indicates that there is an upper limit to the purity under a weak optical pumping field. Experimentally, the impure polarization of light, non-uniform magnetic field, and significant off-resonant transitions affect the state purity. The purities reached up to 98(1)$\%$ in $|1,0\rangle$ and 98(2)$\%$ in $|2,0\rangle$ states after the optimization. The population accumulated in $|1,1\rangle$ state was $96(2)\%$ mainly due to the inhomogeneous magnetic field. 
Using this real-time microwave spectrum, we can dynamically track the population distribution in the optical pumping process.
This technique can significantly impact quantum-information processing in multilevel atomic systems, and our investigation will advance future applications in precision measurement, such as atomic clocks. 

\begin{acknowledgments}
This work was supported by Grants Nos. 109-2112-M-110-008-MY3 and 111-2123-M-006-001 of the National Science and Technology Council, Taiwan. The authors thank Prof. Cheng Chin for the valuable comments on the study.
\end{acknowledgments}



\bibliography{RefMWspectrum}



\setcounter{figure}{0}
\renewcommand{\figurename}{Fig.}
\renewcommand{\thefigure}{S\arabic{figure}}

\end{document}